\documentclass[prl,twocolumn,letterpaper,amssymb,amsmath]{revtex4-1}
\usepackage{times}
\usepackage{graphicx}
\usepackage{amssymb} 
\usepackage{amsmath} 
\usepackage{esvect}
\usepackage{physics}
\usepackage[normalem]{ulem}
\usepackage{booktabs}
\usepackage{tabularx}
\usepackage[english]{babel}
\usepackage{titlesec}
\usepackage{lettrine}
\usepackage{ulem}
\usepackage{soul}
\usepackage[T1]{fontenc}
\titleformat{\section}{\raggedright\bfseries}{\arabic{section}.}{1em}{} 

\usepackage[colorlinks=true, pdfstartview=FitV, linkcolor=blue, citecolor=blue, urlcolor=blue]{hyperref}

\usepackage{afterpage}

\begin{document}

\title{\large Decoding the Stability of Transition-Metal Alloys with Theory-infused Deep Learning}

\author{Yang Huang$^{1,\dagger}$}
\author{Shih-Han Wang$^{1,\dagger}$}
\author{Shuyi Cao$^1$}
\author{Luke E.~K.~Achenie$^1$}
\author{Hongliang Xin$^{1,}$}
\email{hxin@vt.edu}

\affiliation{$^1$Department of Chemical Engineering, Virginia Polytechnic Institute and State University, Blacksburg, VA 24061, USA}

\begin{abstract}
We introduce an interpretable deep learning framework that predicts the cohesive energy of transition-metal alloys (TMAs) by embedding cohesion theory within graph neural networks (GNNs). Beyond accurate prediction of cohesive energy, a key indicator of thermodynamic stability, the model offers mechanistic insights by disentangling energy contributions into physically meaningful components. These data-driven interpretations reveal periodic trends and stability principles governing transition metals. We apply the model to single-atom alloys (SAAs) to assess their thermodynamic resilience against two destabilizing processes: agglomeration (adatom clustering) and segregation (migration into the subsurface). Our analysis shows that these phenomena are governed by distinct physical factors—agglomeration is primarily influenced by localized $d$-orbital coupling, while segregation is dictated by delocalized effects such as wavefunction renormalization. This model thus serves as an explainable AI tool for understanding and guiding the design of stable TMAs, with implications for catalysis and materials discovery.
\end{abstract}

\pacs{31.15.A-,68.35.Ja,82.53.-k,82.20.-w,82.40.-g,78.70.Dm} 
\keywords{deep learning, catalysis, density functional theory, tight-binding theory, alloys} 
\maketitle

\lettrine{T}{ }ransition-metal alloys (TMAs) play pivotal roles in heterogeneous catalysis, offering remarkable activity, selectivity, and tunable functionality across a broad spectrum of reactions~\cite{schwank1985bimetallic,kitchin2004modification,zhang2005controlling}. However, a persistent challenge in TMA design for catalysis is to achieve high stability~\cite{antolini2006stability,bezerra2007review,meier2014design}. For example, Pt-based alloys, though highly active for oxygen reduction, often suffer corrosion and dealloying under operational conditions, leading to leaching of the non-noble component and loss of performance~\cite{zhang2021stabilizing}. Recently, single-atom alloys (SAAs), materials in which isolated metal atoms are atomically dispersed in a host metal or alloy surface, have emerged as a promising new class of TMA catalysts~\cite{gao2024electrifying,zhang2021single,giannakakis2018single,liu2016tackling,zhang2019platinum,huang2024origin}. The unique site geometry of SAAs (with isolated dopant atoms embedded in a host surface) can decouple transition-state energetics from intermediate adsorption, promoting efficient bond activation while minimizing poisoning and even breaking limitations imposed by conventional scaling relations~\cite{lucci2016controlling,reocreux2020controlling,cheng2022single,gao2023synthesis,darby2018carbon,hannagan2020single}. Despite these features, SAAs still face serious stability issues: the dispersed guest atoms can agglomerate into clusters~\cite{mccue2015co,ouyang2021directing} or diffuse into the bulk, leading to the loss of isolated active sites. These challenges underscore the need for easily accessible approaches to predict and understand the thermodynamic stability of alloy catalysts across a vast chemical space.

Recent advances in machine learning (ML) have yielded models that can predict cohesive energies with reasonably high precision~\cite{xie2018crystal,hart2021machine,ubaru2017formation}. However, the opaque nature of most advanced ML models limits their interpretability, hindering the extraction of physicochemical insights for materials design and complicating their extension to unexplored materials.

In this study, we introduce a theory-infused neural network (TinNet) approach for predicting the cohesive energy of TMAs. By integrating cohesion theory within a graph neural network (GNN) framework, our model achieves high accuracy while remaining intrinsically interpretable. Beyond predicting the overall cohesive energy (which reflects the stability of an atomic structure), TinNet provides a breakdown of the contributions from individual physical factors underpinning the cohesion process. This data-driven knowledge clarifies the stability trends observed for transition metals across the periodic table. Furthermore, by applying TinNet to SAAs as a case study, we analyze the relative stability of different atomic configurations. Specifically, we compare monomer versus dimer guest-atom arrangements relevant to agglomeration stability and surface versus subsurface guest placements related to segregation. These two types of atomistic processes are governed by distinct effects, which our model quantifies in terms of localized and delocalized contributions. Overall, TinNet offers a powerful and generalizable tool for understanding the stability mechanisms of alloy materials, particularly for applications in catalysis and materials science.

\begin{figure}[h!]
\centering
\includegraphics[width=0.8\columnwidth]{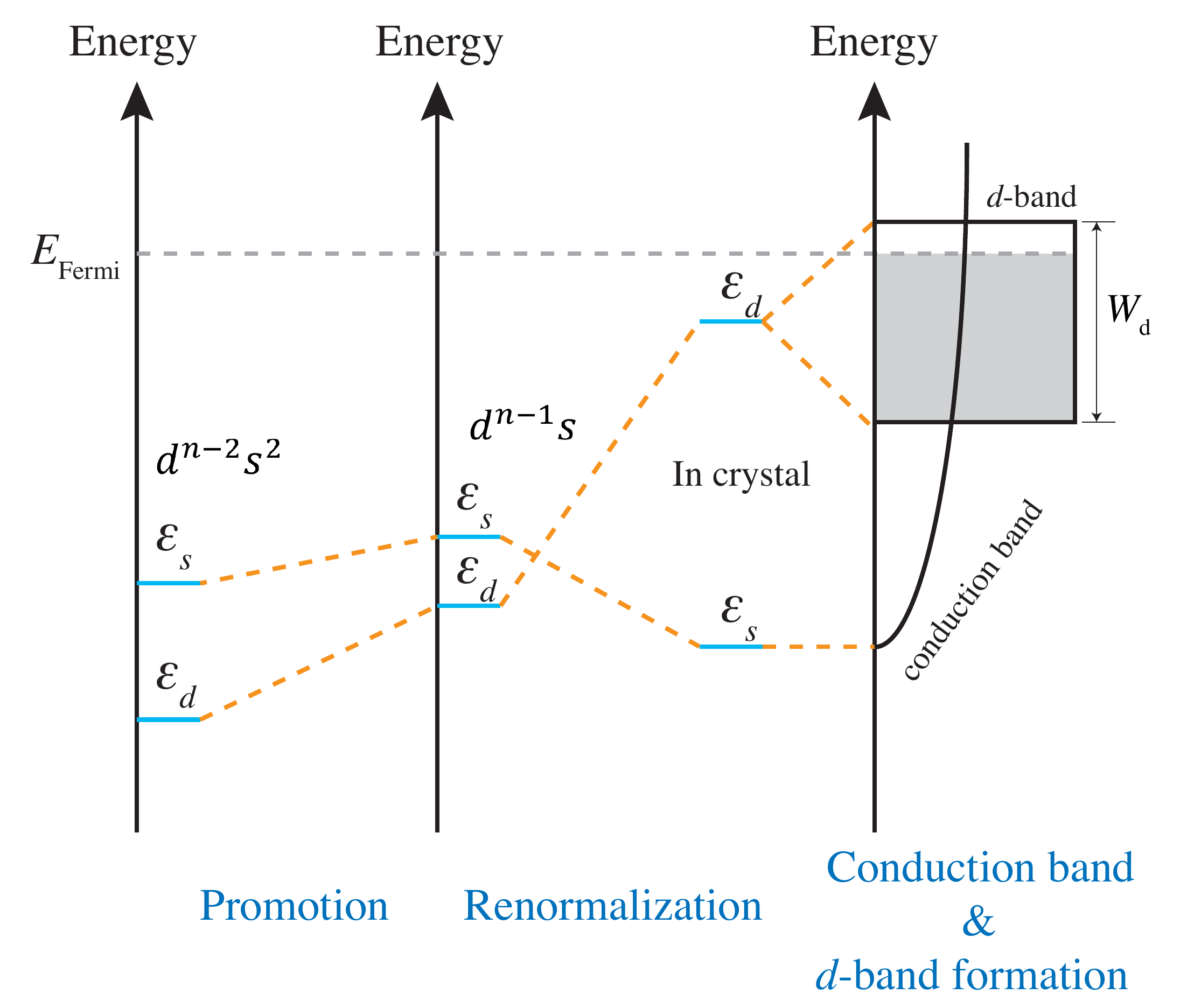}
\caption{Cohesion processes in transition metals as described by the renormalized-atom cohesion theory. Electrons are first promoted from atomic $s$- to $d$-states, followed by renormalization of these states due to interatomic interactions. This leads to the formation of a broad conduction band from delocalized $s$-electrons and a narrower $d$-band from localized
$d$-orbitals.}
\label{fig:Cohesion_theory}
\end{figure}

According to the renormalized-atom cohesion theory (Fig.~\ref{fig:Cohesion_theory}) developed by Watson and co-workers in the 1970s~\cite{hodges1972renormalized,gelatt1977renormalized}, the first step of transition metal cohesion is atomic promotion, in which each free atom is conceptually excited from its ground-state valence configuration, typically $d^{n-2}s^2$, to an excited configuration $d^{n-1}s$, which faciliates bonding in the condensed phase. The corresponding promotion energy for an atom of each element is defined as

\begin{equation}
\label{eq:prom}
\Delta E_{prom} = (d^{n-1}s)_{\text{avg}} - (d^{n-2}s^2)_{\text{min}}
\end{equation} where the subscripts ``avg'' and ``min'' refer to the average and minimum term energies across all multiplets for a given configuration, respectively. These values can be obtained from the NIST Atomic Spectra Database~\cite{ralchenko2005nist}. Notably, for noble metals such as Cu, Ag, and Au, whose ground-state is already in the stable $d^{10}s$ configuration, the promotion energy is effectively zero. The second step is renormalization. In this stage, the excited isolated atoms are compressed into the crystal unit cell. Under the crystal potential, the ``squeezed'' atoms form effective atomic levels, referred to as renormalized states. Related details can be found in the Supplementary Information. For the renormalized $d$-orbital in the crystal potential, the resulting energy level is known as the renormalized $d$-level, denoted $\epsilon_{d,ren}$, which is close to the center of the $d$-band, $\epsilon_d$. Similarly, for orthogonalized plane waves with $k=0$ in the same crystal potential, the energy level is called the renormalized $s$-level, $\epsilon_{s,ren}$, which is close to the bottom of the conduction band, $\epsilon_{\Gamma}$. Based on this, the renormalization energy per atom can be defined as:
\begin{align}
\label{eq:formula}
\Delta E_{\text{ren}} &= N_s\epsilon_{s,\text{ren}} + N_d\epsilon_{d,\text{ren}} - (d^{n-1}s)_{\text{avg}} \nonumber \\
&\approx N_s\epsilon_{\Gamma} + N_d\epsilon_d - (d^{n-1}s)_{\text{avg}}
\end{align} where $N_s$ and $N_d$ are the numbers of valence $s$ and $d$ electrons in their corresponding bands of an atom in the crystal, respectively. The wavefunction renormalization is sensitive to atomic size, crystal volume and the number of $d$-electrons that spatially repel the $s$-electrons. Finally, the renormalized states expand into energy bands due to the Pauli exclusion principle (for delocalized $s$-electrons) and orbital couplings (for localized $d$-electrons). Specifically, based on the free-electron approximation, the conduction band formation energy per atom is given by,
\begin{align}
\label{eq:s-band}
\Delta E_{cond} = \int_{-\infty}^{E_F}\epsilon\rho_s(\epsilon)d\epsilon - N_s\epsilon_{\Gamma} \approx \frac{3\hbar}{10m}(\frac{3\pi^2N_s}{V_{WS}})^{2/3}
\end{align} where $V_{WS}$ is the Wigner-Seitz volume of an atom site, and $m$ is the rest mass of an electron. In the free-electron model, increasing the electron density ($n = \frac{N_s}{V_{WS}}$) raises the Fermi wavevector and thus the Fermi energy, shifting all occupied states to higher kinetic energies. Based on the tight-binding and rectangular band approximations, the $d$-band formation energy is given by,
\begin{align}
\label{eq:d-band}
\Delta E_d 
&= \int_{-\infty}^{E_F} \epsilon\, \rho_d(\epsilon) \, d\epsilon - N_d \epsilon_{d,\text{ren}} \nonumber \\
&\approx \int_{-\infty}^{E_F} (\epsilon - \epsilon_d)\, \rho_d(\epsilon) \, d\epsilon \nonumber \\
&\approx -\frac{W_d}{20} N_d (10 - N_d)
\end{align}
where $W_d$ is the width of an effective rectangular $d$-band whose variance equals that of the original $d$-band. In tight-binding theory, the width of a localized band projected to an atom in the crystal can characterize the overall orbital coupling strengths between that atom and its neighbors. The sum of all these terms: $\Delta E_{prom}$, $\Delta E_{ren}$, $E_{cond}$ and $E_d$ gives the overall cohesive energy for each atom. Among these four contributions, the promotion energy is associated with the atomic electronic configuration prior to crystal formation and is independent of the surrounding chemical environment. The $d$-band formation energy, originating from the coupling of localized $d$-orbitals, is considered a localized effect. In contrast, the renormalization energy and the conduction band formation energy result from the spatial redistribution of electrons in the crystal and are therefore classified as delocalized effects.

The formula based on the renormalized-atom cohesion theory provides a framework for interpreting stability predictions; however, its accuracy can be limited due to the approximations involved. To improve predictive performance, as in Equation \ref{eq:formula}, we parameterized the cohesive energy formula of each atom site and allowed the neural network to learn all relevant physical parameters directly from data. The physical parameters predicted by the neural network include: the number of conduction band electrons $N_s$, the number of $d$-band electrons $N_d$, the $d$-band width $W_d$, and the correction factors $\alpha$ and $\beta$. Details regarding the calculation of DFT-level ground truth values for these physical parameters are provided in the Supplementary Information. It should be noted that $\Delta E_{ren}$ also includes the systematic errors of the single-electron picture behind the renormalized-atom cohesion theory.
\begin{align}
\label{eq:formula}
E_{\text{coh}} = \frac{1}{N} \sum_i^N \bigg[ 
&(d^{n-1}s)_{\text{avg}} - (d^{n-2}s^2)_{\text{min}} + \Delta E_{\text{ren}} \nonumber \\
&+ \frac{3\hbar}{10m} \left( \frac{3\pi^2 N_s}{\beta V_{\text{WS}}} \right)^{2/3} 
- \alpha \frac{W}{20} N_d (10 - N_d) 
\bigg]
\end{align}

\begin{figure}[h!]
\centering
\includegraphics[width=\columnwidth]{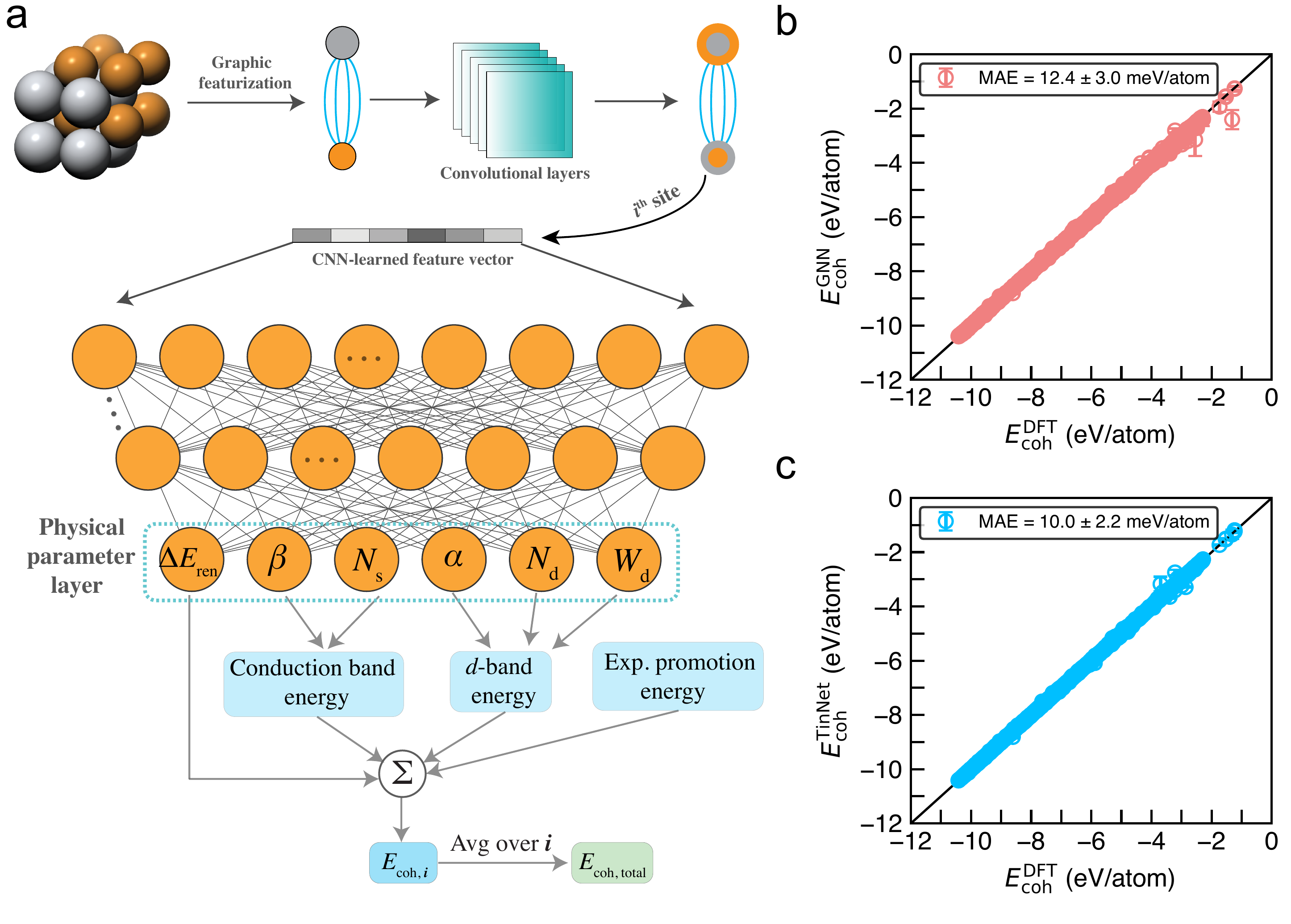}
\caption{(a) The TinNet architecture for accurate and interpretable prediction of cohesive energy. Parity plots of DFT-calculated versus (b) GNN-predicted and (c) TinNet-predicted cohesive energies. Each point represents the average from nested 10-fold cross validation, and error bars denote the standard deviation.}
\label{fig:Ecoh_TinNet}
\end{figure}

The architecture of the theory-infused neural network (TinNet) is shown in Fig. \ref{fig:Ecoh_TinNet}a. The unit cell of the crystal is represented as an undirected graph, serving as the input to a convolutional neural network. A fixed-length feature vector is learned for each atomic site, which is then passed through a fully-connected neural network to predict all the physical parameters and the average total cohesive energy per atom in Equation \ref{eq:formula}. The neural network was trained on a diverse set of transition-metal alloys (see Supplementary Information) to minimize a loss function comprising six terms: the mean absolute errors (MAEs) of $N_s$, $N_d$, $\alpha$, $\beta$ and $W_d$ at the atomic level, and the MAE of $E_{\text{coh}}$ at the alloy level. Although the model explicitly predicts $E_{\text{ren}}$ for each atom, we do not have access to its ground truth with DFT; therefore, the model was supervised using $E_{\text{coh}}$ as the ground truth. Parity plots are shown to compare DFT-calculated and model-predicted $E_{\text{coh}}$ for both the GNN baseline (\ref{fig:Ecoh_TinNet}b) and TinNet (\ref{fig:Ecoh_TinNet}c) across all test sets of 90 models. TinNet achieves a test MAE of 10 meV/atom in cohesive energy prediction, outperforming purely data-driven models such as CGCNN~\cite{xie2018crystal}. In addition, the physical parameters predicted by TinNet are also highly accurate, with MAEs of 0.0056 for $N_s$, 0.0097 for $N_d$, 0.0089 for $\alpha$, 0.0077 for $\beta$, and 0.0550 eV for $W_d$. SAAs with larger unit cell sizes in various dopant-coinage metal combinations were used as unseen systems to test the model's performance. Comparisons between DFT-calculated, GNN-predicted, and TinNet-predicted aggregation and segregation energies are shown in Supplementary Figures~S2 and S3, respectively. The TinNet model demonstrates superior extrapolation capability to these previously unseen systems.

\begin{figure}[h!]
\centering
\includegraphics[width=0.8\columnwidth]{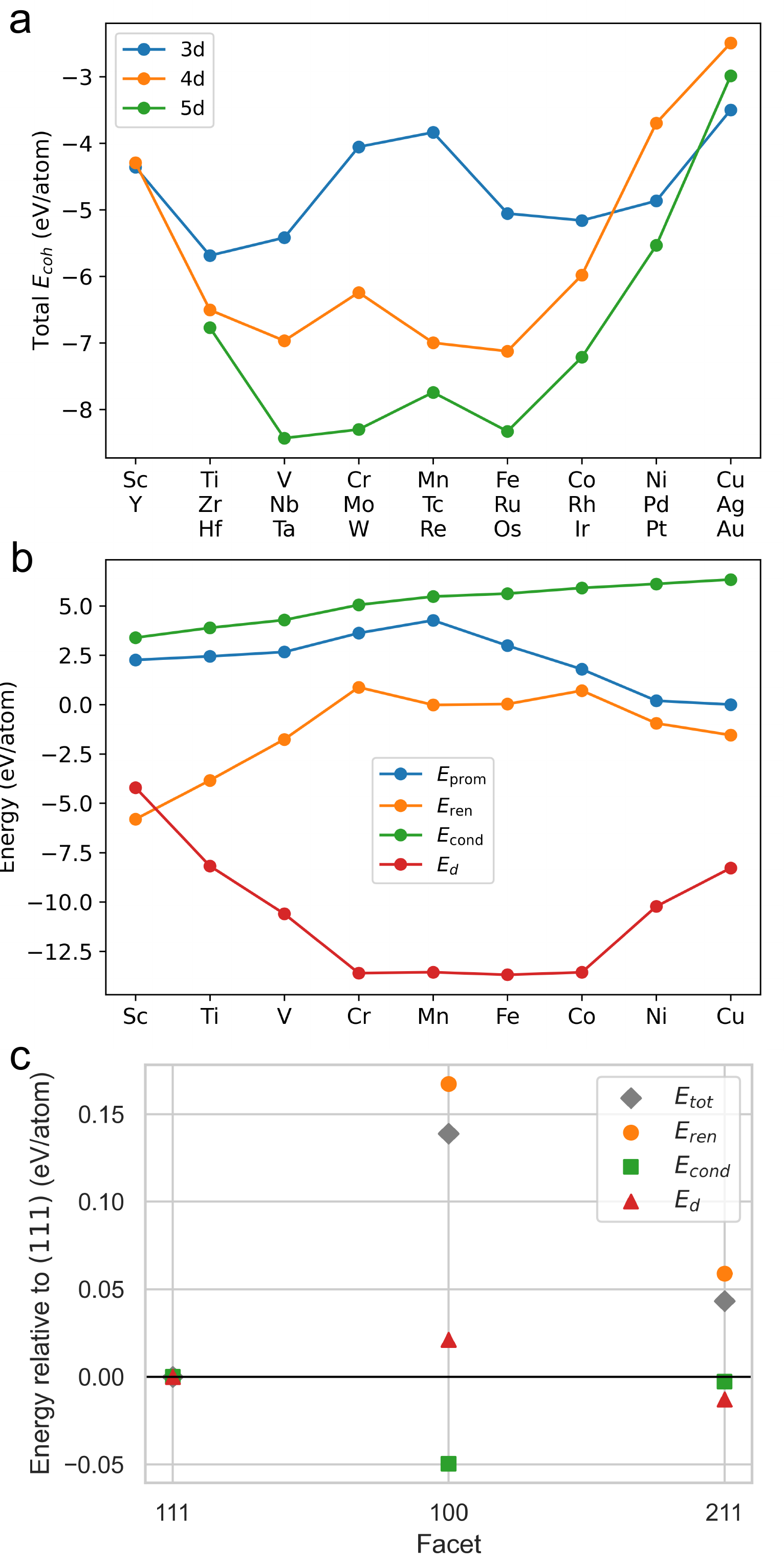}
\caption{TinNet-predicted: (a) total cohesive energy of $3d$, $4d$ and $5d$ metals and (b) contribution components to cohesive energy in $3d$ metals. (c) Cohesive energy and its components of various facets of Pt with different coordination numbers.}
\label{fig:pure_metals}
\end{figure}

Our model not only provides accurate predictions of cohesive energy, but also elucidate the origin of cohesive energy by accurately predicting its contribution terms. For example, as shown in Fig.~\ref{fig:pure_metals}a, the cohesive energies of pure $3d$, $4d$ and $5d$ transition metals exhibit an approximately parabolic trend across each row, with notably higher values near the middle, consistent with experimental observations~\cite{hammer1997theory}. These trends can be interpreted in terms of the individual contribution terms. Taking the 3$d$ metals as an example (Fig.~\ref{fig:pure_metals}b), the dominant contribution comes from the $d$-band formation energy, which itself follows a parabolic trend across the row. Other terms with large variations also vary roughly parabolically, but in the opposite direction, rationalizing the irregularities near the middle of the series. The most significant among these is the promotion energy, which peaks near the center due to the particularly stable ground-state electronic configuration of mid-period elements, and effectively vanishes for the noble metals. The renormalization energy peaks near the center of the transition metal series, where the Wigner–Seitz radius of the metal is smallest relative to the isolated free-atom radius, resulting in maximal renormalization of atomic charge. Due to the obvious large lattice of early transition metals, low conduction band formation energies are observed at the beginning of each row as a result of the small $s$-electron densities. The $d$-band formation energy, which also embeds hybridization between the conduction and $d$ bands, is the largest contributor to the cohesive energy of transition metals. Its nearly parabolic trend across the series is a general feature of the $d$-band contribution. These observations align with prior literatures~\cite{brooks1983exchange,gelatt1977renormalized}. Contribution term trends for the 4$d$ and 5$d$ metals are presented in Fig. S5 and Fig. S6, respectively, in the Supplementary Information.

To understand how atomic coordination environments influence cohesive energy, as shown in Fig.~\ref{fig:pure_metals}c, we analyze the cohesive energy ($E_{\mathrm{coh}}$) and its components across different Pt facets—(111), (100), and (211)—which exhibit distinct surface coordination numbers. For Pt surfaces, the total cohesive energy tends to be more exothermic as the surface coordination number increases, with the (111) facet (CN = 9) being the most stable among the three considered. Although the overall trend is primarily dictated by the renormalization energy ($E_{\mathrm{ren}}$), its physical origin in this context cannot be solely attributed to wavefunction spatial compression, as the crystal structure (fcc) remains unchanged across facets. Instead, the lower $E_{\mathrm{ren}}$ on the (111) surface likely reflects its minimal structural relaxation compared to more open surfaces, where surface atoms tend to reconstruct, incurring additional energetic penalties captured implicitly in $E_{\mathrm{ren}}$. Additionally, the (111) surface retains stronger $d$-orbital overlap among surface atoms due to its dense packing and fewer broken bonds. This is consistent with the observed $d$-band formation energy ($E_d$), which remains lowest for (111) (almost the same as (211)), indicating enhanced $d$–$d$ bonding stabilization. Together, these observations suggest that the high stability of the (111) facet arises not only from its favorable coordination but also from reduced surface relaxation and preserved metallic bonding, all of which are effectively reflected in the decomposed energy components.

\begin{figure}[h!]
\centering
\includegraphics[width=\columnwidth]{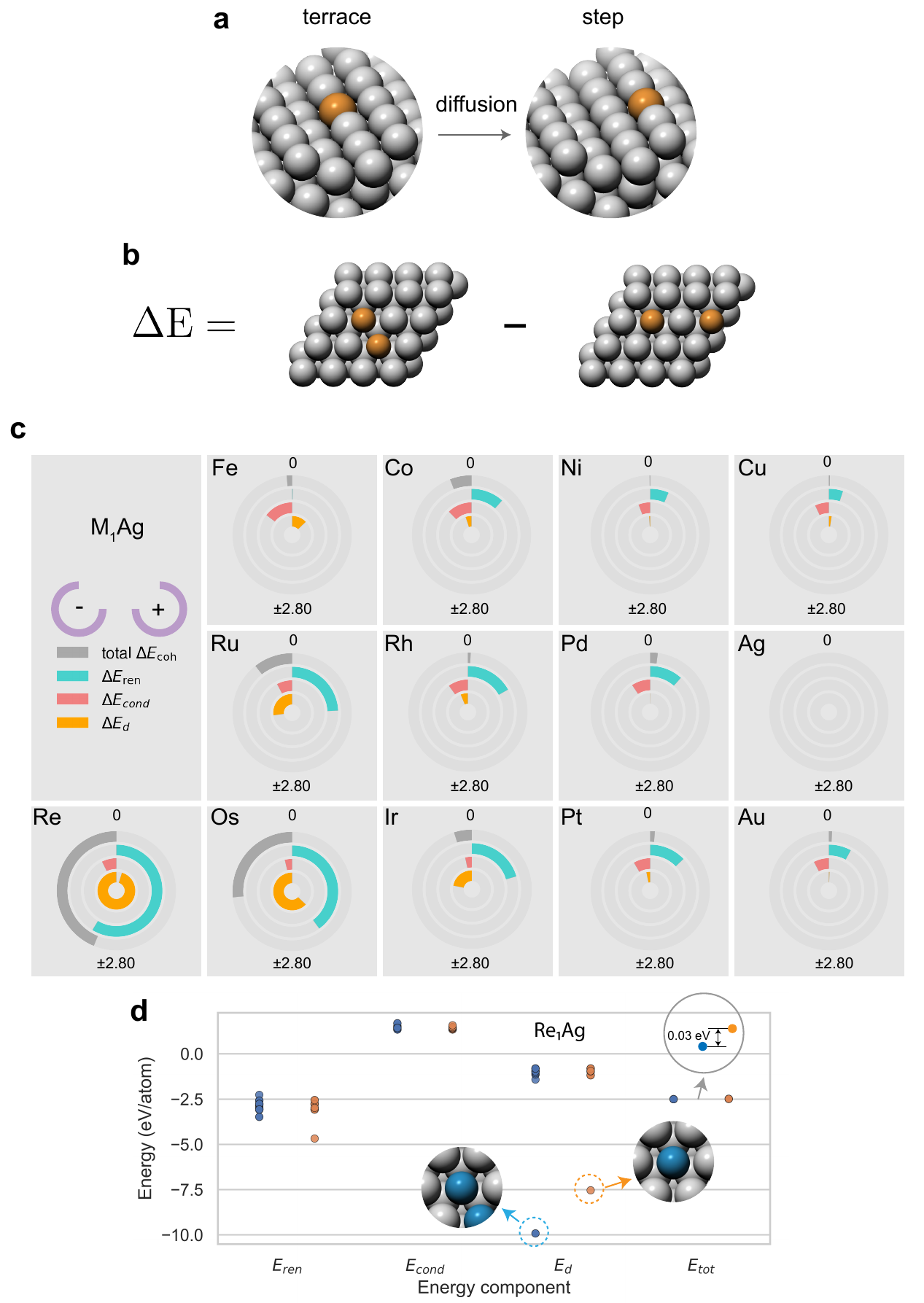}
\caption{(a) Dopant diffusion between terrace and step sites on Ag$_{211}$ surface. (b) Model structures representing dimer-atom and single-atom alloys. The agglomeration energy $\Delta \rm E$ is defined as the total energy difference. (c) Shapley values characterizing the contributions of different terms to the agglomeration energy of the SAAs with Ag as the host. For each SAA, the circles from outer to inner correspond to the total agglomeration energy, renormalization energy difference, conduction band formation energy difference, and $d$-band formation energy difference, respectively. (d) Distributions of each cohesive energy component for atoms in the two Re$_1$Ag configurations. For each energy component, the left distribution in blue is in dimer while the right one in orange is in monomer.}
\label{fig:SAA_monomer_dimer}
\end{figure}

\begin{figure}[h!]
\centering
\includegraphics[width=\columnwidth]{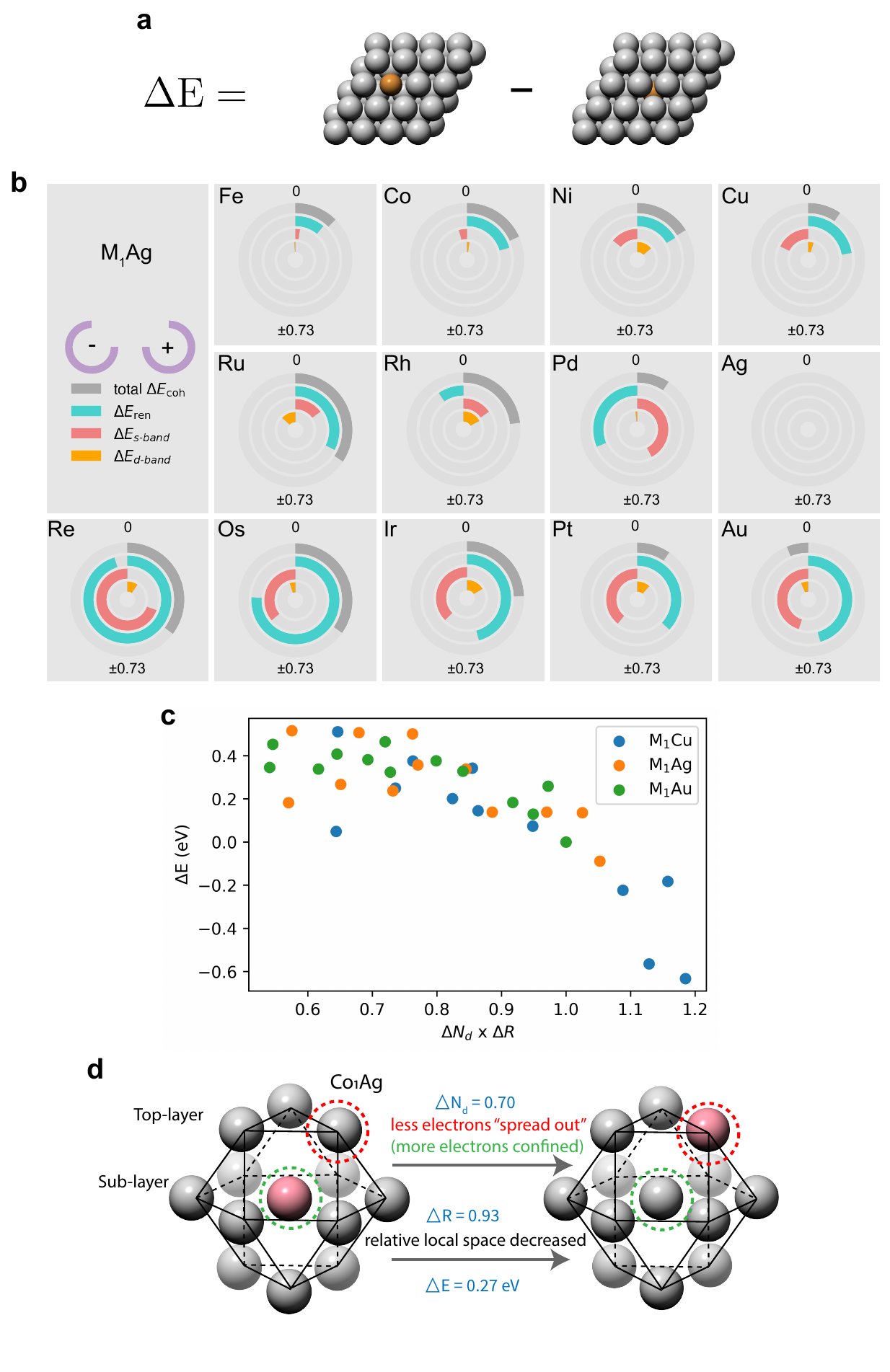}
\caption{(a) Model structures representing SAAs with a guest metal atom located in either the top layer or the sub-layer. The segregation energy $\Delta \rm E$ is defined as the total energy difference between these two configurations. (b) Shapley values characterizing the contributions of different terms to the segregation energy of the SAAs with Ag as the host. For each SAA, the circles from outer to inner correspond to the total segregation energy, renormalization energy difference, conduction band formation energy difference, and $d$-band formation energy difference, respectively. (c) The correlation between a descriptor of delocalized effects and the the segregation energy. The descriptor includes two components: the change in the number of $d$-electrons change at top-layer and relative change in local space volume around the guest atom in the sub-layer, which together characterize the degree of wavefunction renormalization and electron confinement. (d) Mechanisms of the relative stability of the Co$_1$Ag SAA between its top- and sub-layer configurations.}
\label{fig:SAA_1st_2nd_layer}
\end{figure}

Surface mobility sets the stage for SAA chemistry. For example, as shown in Fig.~\ref{fig:SAA_monomer_dimer}a, a dopant that lands on a (111) terrace usually first hop to the under-coordinated step edge, where binding is strongest and catalytic activity is highest. We applied our TinNet model to systematically investigate the stability of SAAs in two representative surface mobility processes: agglomeration (Fig.~\ref{fig:SAA_monomer_dimer}b), which involves in-plane symmetry breaking, and segregation (Fig.~\ref{fig:SAA_1st_2nd_layer}a), which involves out-of-plane symmetry breaking. The guest metals studied include Pt, Pd, Co, Ni, Cu, Os, Ru, Rh, Re, and Ir, while the host metals were Cu, Ag, and Au. To analyze the thermodynamic mechanisms underlying agglomeration, we constructed $4 \times 4 \times 4$ fcc-(111) slabs doped with two guest atoms to represent both configurations: the dimer, where the guest atoms are adjacent, and the monomer, where they are far apart. This consistent setup allows a direct comparison of total energies for the same guest–host combination. The relative stability is defined as the total energy difference between the dimer and monomer configurations and referred to as the agglomeration energy. As shown in Fig.~\ref{fig:SAA_monomer_dimer}c for SAAs with Ag as the host (See Supplementary Information for SAAs with Cu and Au as the host), Shapley values are plotted using the monomer configuration as the baseline to quantify how different energy terms contribute to the agglomeration energy. For each SAA, the circles from outer to inner represent the total agglomeration energy, renormalization energy difference, conduction band formation energy difference, and $d$-band formation energy difference, respectively. In most cases, the absolute agglomeration energies are less than 0.25 eV, indicating a weak thermodynamic driven force. However, when Re or Os is used as the guest metal, the agglomeration energy becomes highly negative, suggesting a strong thermodynamic preference for the dimer configuration. In these cases, the $d$-band formation energy is clearly the dominant factor. The structural difference between the dimer and monomer lies in their local bonding environment: guest–host bonds dominate in the monomer configuration, while guest–guest bonds appear in the dimer. According to tight-binding theory, the $d$-band width $W$ of an atom site in Eqn.~\ref{eq:d-band} is proportional to the total orbital coupling strengths~\cite{huang2024unifying,kitchin2004role} between that site and its neighboring atoms. The coupling strength of each pair of atoms is positively correlated with the $d$-orbital radius $r_d$ of the neighboring atom. Therefore, for a guest metal site in both configurations, their $d$-band formation energy difference can be approximated as $-N_d^{g} (10-N_d^{g})r_d^g (r_d^g - r_d^{h})$, where $g$ and $h$ denote the guest and host metals, respectively. This descriptor becomes most negative when $N_d^g$ approaches to 5 and $r_d^g$ is in its maximum while greater than $r_d^h$. This scenario is realized by Re ($N_d=5$) and Os ($N_d=6$), both of which have the largest $r_d$ among the late transition metals considered as guest elements. A higher $r_d$ of the guest metal indicates stronger guest-guest bonding and a stronger tendency toward dimer formation. A site-wise analysis was performed on Re$_1$Ag(111) system, as shown in Fig.~\ref{fig:SAA_monomer_dimer}d, where the contributions to cohesive energy from each atom site are plotted. For each energy component, the left distribution in blue is in dimer while the right one in orange is in monomer. The results show that only the $d$-band formation energy difference at the Re site aligns in sign with the total cohesive energy difference, confirming its dominant role in agglomeration. The enhanced stability of the dimer configuration is attributed to stronger Re–Re than Re–Ag $d$-orbital coupling, due to the significantly larger $d$-orbital radius of Re (0.88) compared to Ag (0.66), which substantially lowers the $d$-band formation energy at Re sites.

For SAAs with guest atoms positioned in either the top or sub-layer, the relative stability is defined as the total energy difference between the two configurations, referred to as the segregation energy. As shown in Fig.~\ref{fig:SAA_1st_2nd_layer}b for SAAs with Ag as the host (See Supplementary Information for SAAs with Cu and Au as the host), in most cases, the sub-layer configuration tends to be more stable with non-negligible segregation energies. Among the three contribution energy terms, the $d$-band formation energy can generally be neglected in this context because there is no significant change in bond types during segregation. In our calculations, the renormalization energy absorbs all systematic and numerical errors, making both its sign and magnitude unreliable. As a result, it is more meaningful to combine the renormalization and conduction band formation energies—both of which are volume-dependent, delocalized effects into a single analytical quantity. Two key distinctions exist between the top- and sub-layer configurations. First, the degree of electron confinement differs: a guest atom in the top layer experiences less confinement, while one in the sub-layer is more embedded in the bulk environment. Second, out-of-plane symmetry breaking in segregation can lead to larger variations in available free space than in-plane symmetry breaking during agglomeration. To quantify the energetic implications of these effects, we define a simple descriptor, $\Delta N_d \cdot \Delta R$. Here, $\Delta N_d$ is the ratio of $d$-electron count between guest and host atoms (i.e., $N_d^{\text{guest}} / N_d^{\text{host}}$), which serves as a proxy for electronic confinement. A smaller $\Delta N_d$ implies fewer guest $d$-electrons can delocalize at the surface, leading to stronger confinement in the bulk and, consequently, stronger $d$–$s$ repulsion and higher renormalization energy. $\Delta R$ represents the ratio of relative local space radius $R$ between the two configurations. The relative local space radius $R$ of each configuration is defined as $R^{\text{crystal}}/R^{\text{atom}}$, where $R^{\text{crystal}}$ is the radius of the coordination polyhedron centered at the guest atom in the sub-layer (based on its 12 nearest neighbors), and $R^{\text{atom}}$ is the free atom radius of the guest element~\cite{clementi1963atomic}. This geometric measure captures both the available spatial volume for electron delocalization, which influences the kinetic energy of conduction electrons, and the degree of wavefunction confinement, which affects renormalization energy. In short, a lower $\Delta R$ increases both renormalization and $s$-band formation energies. As shown in Fig.\ref{fig:SAA_1st_2nd_layer}c, this composite descriptor qualitatively explains the observed trends in segregation energy across different SAAs systems. A local case study on Co$_1$Ag(111), as illustrated in Fig.\ref{fig:SAA_1st_2nd_layer}d, supports this interpretation: the sub-layer configuration is more stable due to less confined $d$-electrons and a larger relative local space, both of which reduce the total cohesive energy. As noted here, in most systems studied, the sub-layer configuration is thermodynamically favored. This trend is consistent with the fact that guest metals typically have fewer $d$-electrons than the host metals Cu, Ag, and Au, which all have filled $d^{10}$ shells. When the guest atom occupies the sub-layer and the host atom remains at the surface, more electrons are distributed near the surface and fewer are confined within the bulk, resulting in a lower overall cohesive energy.

It is important to note that all of the above analyses are based solely on thermodynamic preferences. However, actual surface configurations are determined by both thermodynamics and kinetics. If the system is kinetically stable, meaning that there are significant energy barriers along the atomic diffusion pathways, then even transformations that are thermodynamically favorable, such as monomer-to-dimer agglomeration or top-to-sub-layer segregation, may not occur in practice. Furthermore, our current study focuses exclusively on bare surfaces and does not account for adsorbate-induced effects, which can be substantial under realistic reaction conditions and may influence the preferred configurations of surface atoms.

While this work focuses on applying TinNet to the stability analysis of single-atom alloys (SAAs), the framework’s potential extends well beyond this domain. By integrating general physical theory with graph-based machine learning, the model is particularly well-suited for application to more complex compositional spaces, such as high-entropy alloys (HEAs)~\cite{bueno2022quinary,huang2024unraveling} and doped dimer or trimer SAA clusters~\cite{kress2023priori}, where accurate stability prediction remains a significant challenge. These systems involve combinatorially large design spaces and complex bonding environments, making them ideal for TinNet-based analysis. In future work, we plan to expand in these directions and further enhance TinNet to support the rational discovery and design of next-generation alloy catalysts that combine high stability with tailored functional properties.

In conclusion, we introduce TinNet, an interpretable graph neural network model that integrates cohesion theory into deep learning for accurate and physically grounded prediction of cohesive energy in transition metal alloys (TMAs). TinNet not only captures overall energy trends but also reveals the underlying contributions from both localized and delocalized interactions. When applied to single-atom alloys (SAAs), the model helps elucidate the stability mechanisms governing agglomeration and segregation, highlighting the interplay between electronic structure and spatial confinement in determining thermodynamic preferences. While this study presents an initial demonstration of TinNet’s utility for stability analysis, we believe the framework holds broader potential for exploring more complex alloy systems with larger and more diverse design spaces. TinNet thus establishes a promising foundation for interpretable and scalable materials discovery.

\bibliography{references.bib}

\begin{thebibliography}{35}%
\makeatletter
\providecommand \@ifxundefined [1]{%
 \@ifx{#1\undefined}
}%
\providecommand \@ifnum [1]{%
 \ifnum #1\expandafter \@firstoftwo
 \else \expandafter \@secondoftwo
 \fi
}%
\providecommand \@ifx [1]{%
 \ifx #1\expandafter \@firstoftwo
 \else \expandafter \@secondoftwo
 \fi
}%
\providecommand \natexlab [1]{#1}%
\providecommand \enquote  [1]{``#1''}%
\providecommand \bibnamefont  [1]{#1}%
\providecommand \bibfnamefont [1]{#1}%
\providecommand \citenamefont [1]{#1}%
\providecommand \href@noop [0]{\@secondoftwo}%
\providecommand \href [0]{\begingroup \@sanitize@url \@href}%
\providecommand \@href[1]{\@@startlink{#1}\@@href}%
\providecommand \@@href[1]{\endgroup#1\@@endlink}%
\providecommand \@sanitize@url [0]{\catcode `\\12\catcode `\$12\catcode `\&12\catcode `\#12\catcode `\^12\catcode `\_12\catcode `\%12\relax}%
\providecommand \@@startlink[1]{}%
\providecommand \@@endlink[0]{}%
\providecommand \url  [0]{\begingroup\@sanitize@url \@url }%
\providecommand \@url [1]{\endgroup\@href {#1}{\urlprefix }}%
\providecommand \urlprefix  [0]{URL }%
\providecommand \Eprint [0]{\href }%
\providecommand \doibase [0]{http://dx.doi.org/}%
\providecommand \selectlanguage [0]{\@gobble}%
\providecommand \bibinfo  [0]{\@secondoftwo}%
\providecommand \bibfield  [0]{\@secondoftwo}%
\providecommand \translation [1]{[#1]}%
\providecommand \BibitemOpen [0]{}%
\providecommand \bibitemStop [0]{}%
\providecommand \bibitemNoStop [0]{.\EOS\space}%
\providecommand \EOS [0]{\spacefactor3000\relax}%
\providecommand \BibitemShut  [1]{\csname bibitem#1\endcsname}%
\let\auto@bib@innerbib\@empty
\bibitem [{\citenamefont {Schwank}(1985)}]{schwank1985bimetallic}%
  \BibitemOpen
  \bibfield  {author} {\bibinfo {author} {\bibfnamefont {J.~W.}\ \bibnamefont {Schwank}},\ }\href@noop {} {\  (\bibinfo {year} {1985})}\BibitemShut {NoStop}%
\bibitem [{\citenamefont {Kitchin}\ \emph {et~al.}(2004{\natexlab{a}})\citenamefont {Kitchin}, \citenamefont {N{\o}rskov}, \citenamefont {Barteau},\ and\ \citenamefont {Chen}}]{kitchin2004modification}%
  \BibitemOpen
  \bibfield  {author} {\bibinfo {author} {\bibfnamefont {J.}~\bibnamefont {Kitchin}}, \bibinfo {author} {\bibfnamefont {J.~K.}\ \bibnamefont {N{\o}rskov}}, \bibinfo {author} {\bibfnamefont {M.}~\bibnamefont {Barteau}}, \ and\ \bibinfo {author} {\bibfnamefont {J.}~\bibnamefont {Chen}},\ }\href@noop {} {\bibfield  {journal} {\bibinfo  {journal} {The Journal of chemical physics}\ }\textbf {\bibinfo {volume} {120}},\ \bibinfo {pages} {10240} (\bibinfo {year} {2004}{\natexlab{a}})}\BibitemShut {NoStop}%
\bibitem [{\citenamefont {Zhang}\ \emph {et~al.}(2005)\citenamefont {Zhang}, \citenamefont {Vukmirovic}, \citenamefont {Xu}, \citenamefont {Mavrikakis},\ and\ \citenamefont {Adzic}}]{zhang2005controlling}%
  \BibitemOpen
  \bibfield  {author} {\bibinfo {author} {\bibfnamefont {J.}~\bibnamefont {Zhang}}, \bibinfo {author} {\bibfnamefont {M.~B.}\ \bibnamefont {Vukmirovic}}, \bibinfo {author} {\bibfnamefont {Y.}~\bibnamefont {Xu}}, \bibinfo {author} {\bibfnamefont {M.}~\bibnamefont {Mavrikakis}}, \ and\ \bibinfo {author} {\bibfnamefont {R.~R.}\ \bibnamefont {Adzic}},\ }\href@noop {} {\bibfield  {journal} {\bibinfo  {journal} {Angewandte Chemie}\ }\textbf {\bibinfo {volume} {117}},\ \bibinfo {pages} {2170} (\bibinfo {year} {2005})}\BibitemShut {NoStop}%
\bibitem [{\citenamefont {Antolini}\ \emph {et~al.}(2006)\citenamefont {Antolini}, \citenamefont {Salgado},\ and\ \citenamefont {Gonzalez}}]{antolini2006stability}%
  \BibitemOpen
  \bibfield  {author} {\bibinfo {author} {\bibfnamefont {E.}~\bibnamefont {Antolini}}, \bibinfo {author} {\bibfnamefont {J.~R.}\ \bibnamefont {Salgado}}, \ and\ \bibinfo {author} {\bibfnamefont {E.~R.}\ \bibnamefont {Gonzalez}},\ }\href@noop {} {\bibfield  {journal} {\bibinfo  {journal} {Journal of Power Sources}\ }\textbf {\bibinfo {volume} {160}},\ \bibinfo {pages} {957} (\bibinfo {year} {2006})}\BibitemShut {NoStop}%
\bibitem [{\citenamefont {Bezerra}\ \emph {et~al.}(2007)\citenamefont {Bezerra}, \citenamefont {Zhang}, \citenamefont {Liu}, \citenamefont {Lee}, \citenamefont {Marques}, \citenamefont {Marques}, \citenamefont {Wang},\ and\ \citenamefont {Zhang}}]{bezerra2007review}%
  \BibitemOpen
  \bibfield  {author} {\bibinfo {author} {\bibfnamefont {C.~W.}\ \bibnamefont {Bezerra}}, \bibinfo {author} {\bibfnamefont {L.}~\bibnamefont {Zhang}}, \bibinfo {author} {\bibfnamefont {H.}~\bibnamefont {Liu}}, \bibinfo {author} {\bibfnamefont {K.}~\bibnamefont {Lee}}, \bibinfo {author} {\bibfnamefont {A.~L.}\ \bibnamefont {Marques}}, \bibinfo {author} {\bibfnamefont {E.~P.}\ \bibnamefont {Marques}}, \bibinfo {author} {\bibfnamefont {H.}~\bibnamefont {Wang}}, \ and\ \bibinfo {author} {\bibfnamefont {J.}~\bibnamefont {Zhang}},\ }\href@noop {} {\bibfield  {journal} {\bibinfo  {journal} {Journal of Power Sources}\ }\textbf {\bibinfo {volume} {173}},\ \bibinfo {pages} {891} (\bibinfo {year} {2007})}\BibitemShut {NoStop}%
\bibitem [{\citenamefont {Meier}\ \emph {et~al.}(2014)\citenamefont {Meier}, \citenamefont {Galeano}, \citenamefont {Katsounaros}, \citenamefont {Witte}, \citenamefont {Bongard}, \citenamefont {Topalov}, \citenamefont {Baldizzone}, \citenamefont {Mezzavilla}, \citenamefont {Sch{\"u}th},\ and\ \citenamefont {Mayrhofer}}]{meier2014design}%
  \BibitemOpen
  \bibfield  {author} {\bibinfo {author} {\bibfnamefont {J.~C.}\ \bibnamefont {Meier}}, \bibinfo {author} {\bibfnamefont {C.}~\bibnamefont {Galeano}}, \bibinfo {author} {\bibfnamefont {I.}~\bibnamefont {Katsounaros}}, \bibinfo {author} {\bibfnamefont {J.}~\bibnamefont {Witte}}, \bibinfo {author} {\bibfnamefont {H.~J.}\ \bibnamefont {Bongard}}, \bibinfo {author} {\bibfnamefont {A.~A.}\ \bibnamefont {Topalov}}, \bibinfo {author} {\bibfnamefont {C.}~\bibnamefont {Baldizzone}}, \bibinfo {author} {\bibfnamefont {S.}~\bibnamefont {Mezzavilla}}, \bibinfo {author} {\bibfnamefont {F.}~\bibnamefont {Sch{\"u}th}}, \ and\ \bibinfo {author} {\bibfnamefont {K.~J.}\ \bibnamefont {Mayrhofer}},\ }\href@noop {} {\bibfield  {journal} {\bibinfo  {journal} {Beilstein journal of nanotechnology}\ }\textbf {\bibinfo {volume} {5}},\ \bibinfo {pages} {44} (\bibinfo {year} {2014})}\BibitemShut {NoStop}%
\bibitem [{\citenamefont {Zhang}\ \emph {et~al.}(2021{\natexlab{a}})\citenamefont {Zhang}, \citenamefont {Yuan}, \citenamefont {Gao}, \citenamefont {Zeng}, \citenamefont {Li},\ and\ \citenamefont {Huang}}]{zhang2021stabilizing}%
  \BibitemOpen
  \bibfield  {author} {\bibinfo {author} {\bibfnamefont {J.}~\bibnamefont {Zhang}}, \bibinfo {author} {\bibfnamefont {Y.}~\bibnamefont {Yuan}}, \bibinfo {author} {\bibfnamefont {L.}~\bibnamefont {Gao}}, \bibinfo {author} {\bibfnamefont {G.}~\bibnamefont {Zeng}}, \bibinfo {author} {\bibfnamefont {M.}~\bibnamefont {Li}}, \ and\ \bibinfo {author} {\bibfnamefont {H.}~\bibnamefont {Huang}},\ }\href@noop {} {\bibfield  {journal} {\bibinfo  {journal} {Advanced Materials}\ }\textbf {\bibinfo {volume} {33}},\ \bibinfo {pages} {2006494} (\bibinfo {year} {2021}{\natexlab{a}})}\BibitemShut {NoStop}%
\bibitem [{\citenamefont {Gao}\ \emph {et~al.}(2024)\citenamefont {Gao}, \citenamefont {Han}, \citenamefont {Liu},\ and\ \citenamefont {Zhu}}]{gao2024electrifying}%
  \BibitemOpen
  \bibfield  {author} {\bibinfo {author} {\bibfnamefont {Q.}~\bibnamefont {Gao}}, \bibinfo {author} {\bibfnamefont {X.}~\bibnamefont {Han}}, \bibinfo {author} {\bibfnamefont {Y.}~\bibnamefont {Liu}}, \ and\ \bibinfo {author} {\bibfnamefont {H.}~\bibnamefont {Zhu}},\ }\href@noop {} {\bibfield  {journal} {\bibinfo  {journal} {ACS catalysis}\ }\textbf {\bibinfo {volume} {14}},\ \bibinfo {pages} {6045} (\bibinfo {year} {2024})}\BibitemShut {NoStop}%
\bibitem [{\citenamefont {Zhang}\ \emph {et~al.}(2021{\natexlab{b}})\citenamefont {Zhang}, \citenamefont {Walsh}, \citenamefont {Yu},\ and\ \citenamefont {Zhang}}]{zhang2021single}%
  \BibitemOpen
  \bibfield  {author} {\bibinfo {author} {\bibfnamefont {T.}~\bibnamefont {Zhang}}, \bibinfo {author} {\bibfnamefont {A.~G.}\ \bibnamefont {Walsh}}, \bibinfo {author} {\bibfnamefont {J.}~\bibnamefont {Yu}}, \ and\ \bibinfo {author} {\bibfnamefont {P.}~\bibnamefont {Zhang}},\ }\href@noop {} {\bibfield  {journal} {\bibinfo  {journal} {Chemical Society Reviews}\ }\textbf {\bibinfo {volume} {50}},\ \bibinfo {pages} {569} (\bibinfo {year} {2021}{\natexlab{b}})}\BibitemShut {NoStop}%
\bibitem [{\citenamefont {Giannakakis}\ \emph {et~al.}(2018)\citenamefont {Giannakakis}, \citenamefont {Flytzani-Stephanopoulos},\ and\ \citenamefont {Sykes}}]{giannakakis2018single}%
  \BibitemOpen
  \bibfield  {author} {\bibinfo {author} {\bibfnamefont {G.}~\bibnamefont {Giannakakis}}, \bibinfo {author} {\bibfnamefont {M.}~\bibnamefont {Flytzani-Stephanopoulos}}, \ and\ \bibinfo {author} {\bibfnamefont {E.~C.~H.}\ \bibnamefont {Sykes}},\ }\href@noop {} {\bibfield  {journal} {\bibinfo  {journal} {Accounts of chemical research}\ }\textbf {\bibinfo {volume} {52}},\ \bibinfo {pages} {237} (\bibinfo {year} {2018})}\BibitemShut {NoStop}%
\bibitem [{\citenamefont {Liu}\ \emph {et~al.}(2016)\citenamefont {Liu}, \citenamefont {Lucci}, \citenamefont {Yang}, \citenamefont {Lee}, \citenamefont {Marcinkowski}, \citenamefont {Therrien}, \citenamefont {Williams}, \citenamefont {Sykes},\ and\ \citenamefont {Flytzani-Stephanopoulos}}]{liu2016tackling}%
  \BibitemOpen
  \bibfield  {author} {\bibinfo {author} {\bibfnamefont {J.}~\bibnamefont {Liu}}, \bibinfo {author} {\bibfnamefont {F.~R.}\ \bibnamefont {Lucci}}, \bibinfo {author} {\bibfnamefont {M.}~\bibnamefont {Yang}}, \bibinfo {author} {\bibfnamefont {S.}~\bibnamefont {Lee}}, \bibinfo {author} {\bibfnamefont {M.~D.}\ \bibnamefont {Marcinkowski}}, \bibinfo {author} {\bibfnamefont {A.~J.}\ \bibnamefont {Therrien}}, \bibinfo {author} {\bibfnamefont {C.~T.}\ \bibnamefont {Williams}}, \bibinfo {author} {\bibfnamefont {E.~C.~H.}\ \bibnamefont {Sykes}}, \ and\ \bibinfo {author} {\bibfnamefont {M.}~\bibnamefont {Flytzani-Stephanopoulos}},\ }\href@noop {} {\bibfield  {journal} {\bibinfo  {journal} {Journal of the American Chemical Society}\ }\textbf {\bibinfo {volume} {138}},\ \bibinfo {pages} {6396} (\bibinfo {year} {2016})}\BibitemShut {NoStop}%
\bibitem [{\citenamefont {Zhang}\ \emph {et~al.}(2019)\citenamefont {Zhang}, \citenamefont {Cui}, \citenamefont {Feng}, \citenamefont {Chen}, \citenamefont {Wang}, \citenamefont {Wang}, \citenamefont {Zhang}, \citenamefont {Zheng}, \citenamefont {Hong},\ and\ \citenamefont {Wei}}]{zhang2019platinum}%
  \BibitemOpen
  \bibfield  {author} {\bibinfo {author} {\bibfnamefont {X.}~\bibnamefont {Zhang}}, \bibinfo {author} {\bibfnamefont {G.}~\bibnamefont {Cui}}, \bibinfo {author} {\bibfnamefont {H.}~\bibnamefont {Feng}}, \bibinfo {author} {\bibfnamefont {L.}~\bibnamefont {Chen}}, \bibinfo {author} {\bibfnamefont {H.}~\bibnamefont {Wang}}, \bibinfo {author} {\bibfnamefont {B.}~\bibnamefont {Wang}}, \bibinfo {author} {\bibfnamefont {X.}~\bibnamefont {Zhang}}, \bibinfo {author} {\bibfnamefont {L.}~\bibnamefont {Zheng}}, \bibinfo {author} {\bibfnamefont {S.}~\bibnamefont {Hong}}, \ and\ \bibinfo {author} {\bibfnamefont {M.}~\bibnamefont {Wei}},\ }\href@noop {} {\bibfield  {journal} {\bibinfo  {journal} {Nature Communications}\ }\textbf {\bibinfo {volume} {10}},\ \bibinfo {pages} {5812} (\bibinfo {year} {2019})}\BibitemShut {NoStop}%
\bibitem [{\citenamefont {Huang}\ \emph {et~al.}(2024{\natexlab{a}})\citenamefont {Huang}, \citenamefont {Wang}, \citenamefont {Achenie}, \citenamefont {Choudhary},\ and\ \citenamefont {Xin}}]{huang2024origin}%
  \BibitemOpen
  \bibfield  {author} {\bibinfo {author} {\bibfnamefont {Y.}~\bibnamefont {Huang}}, \bibinfo {author} {\bibfnamefont {S.-H.}\ \bibnamefont {Wang}}, \bibinfo {author} {\bibfnamefont {L.~E.}\ \bibnamefont {Achenie}}, \bibinfo {author} {\bibfnamefont {K.}~\bibnamefont {Choudhary}}, \ and\ \bibinfo {author} {\bibfnamefont {H.}~\bibnamefont {Xin}},\ }\href@noop {} {\bibfield  {journal} {\bibinfo  {journal} {The Journal of Chemical Physics}\ }\textbf {\bibinfo {volume} {161}} (\bibinfo {year} {2024}{\natexlab{a}})}\BibitemShut {NoStop}%
\bibitem [{\citenamefont {Lucci}\ \emph {et~al.}(2016)\citenamefont {Lucci}, \citenamefont {Darby}, \citenamefont {Mattera}, \citenamefont {Ivimey}, \citenamefont {Therrien}, \citenamefont {Michaelides}, \citenamefont {Stamatakis},\ and\ \citenamefont {Sykes}}]{lucci2016controlling}%
  \BibitemOpen
  \bibfield  {author} {\bibinfo {author} {\bibfnamefont {F.~R.}\ \bibnamefont {Lucci}}, \bibinfo {author} {\bibfnamefont {M.~T.}\ \bibnamefont {Darby}}, \bibinfo {author} {\bibfnamefont {M.~F.}\ \bibnamefont {Mattera}}, \bibinfo {author} {\bibfnamefont {C.~J.}\ \bibnamefont {Ivimey}}, \bibinfo {author} {\bibfnamefont {A.~J.}\ \bibnamefont {Therrien}}, \bibinfo {author} {\bibfnamefont {A.}~\bibnamefont {Michaelides}}, \bibinfo {author} {\bibfnamefont {M.}~\bibnamefont {Stamatakis}}, \ and\ \bibinfo {author} {\bibfnamefont {E.~C.~H.}\ \bibnamefont {Sykes}},\ }\href@noop {} {\bibfield  {journal} {\bibinfo  {journal} {The journal of physical chemistry letters}\ }\textbf {\bibinfo {volume} {7}},\ \bibinfo {pages} {480} (\bibinfo {year} {2016})}\BibitemShut {NoStop}%
\bibitem [{\citenamefont {R{\'e}ocreux}\ \emph {et~al.}(2020)\citenamefont {R{\'e}ocreux}, \citenamefont {Kress}, \citenamefont {Hannagan}, \citenamefont {{\c{C}}{\i}nar}, \citenamefont {Stamatakis},\ and\ \citenamefont {Sykes}}]{reocreux2020controlling}%
  \BibitemOpen
  \bibfield  {author} {\bibinfo {author} {\bibfnamefont {R.}~\bibnamefont {R{\'e}ocreux}}, \bibinfo {author} {\bibfnamefont {P.~L.}\ \bibnamefont {Kress}}, \bibinfo {author} {\bibfnamefont {R.~T.}\ \bibnamefont {Hannagan}}, \bibinfo {author} {\bibfnamefont {V.}~\bibnamefont {{\c{C}}{\i}nar}}, \bibinfo {author} {\bibfnamefont {M.}~\bibnamefont {Stamatakis}}, \ and\ \bibinfo {author} {\bibfnamefont {E.~C.~H.}\ \bibnamefont {Sykes}},\ }\href@noop {} {\bibfield  {journal} {\bibinfo  {journal} {The Journal of Physical Chemistry Letters}\ }\textbf {\bibinfo {volume} {11}},\ \bibinfo {pages} {8751} (\bibinfo {year} {2020})}\BibitemShut {NoStop}%
\bibitem [{\citenamefont {Cheng}\ \emph {et~al.}(2022)\citenamefont {Cheng}, \citenamefont {Wang}, \citenamefont {Lu}, \citenamefont {Zheng}, \citenamefont {Sun}, \citenamefont {Li}, \citenamefont {Chen},\ and\ \citenamefont {Zhang}}]{cheng2022single}%
  \BibitemOpen
  \bibfield  {author} {\bibinfo {author} {\bibfnamefont {X.}~\bibnamefont {Cheng}}, \bibinfo {author} {\bibfnamefont {Y.}~\bibnamefont {Wang}}, \bibinfo {author} {\bibfnamefont {Y.}~\bibnamefont {Lu}}, \bibinfo {author} {\bibfnamefont {L.}~\bibnamefont {Zheng}}, \bibinfo {author} {\bibfnamefont {S.}~\bibnamefont {Sun}}, \bibinfo {author} {\bibfnamefont {H.}~\bibnamefont {Li}}, \bibinfo {author} {\bibfnamefont {G.}~\bibnamefont {Chen}}, \ and\ \bibinfo {author} {\bibfnamefont {J.}~\bibnamefont {Zhang}},\ }\href@noop {} {\bibfield  {journal} {\bibinfo  {journal} {Applied Catalysis B: Environmental}\ }\textbf {\bibinfo {volume} {306}},\ \bibinfo {pages} {121112} (\bibinfo {year} {2022})}\BibitemShut {NoStop}%
\bibitem [{\citenamefont {Gao}\ \emph {et~al.}(2023)\citenamefont {Gao}, \citenamefont {Yao}, \citenamefont {Pillai}, \citenamefont {Zang}, \citenamefont {Han}, \citenamefont {Liu}, \citenamefont {Yu}, \citenamefont {Yan}, \citenamefont {Min}, \citenamefont {Zhang} \emph {et~al.}}]{gao2023synthesis}%
  \BibitemOpen
  \bibfield  {author} {\bibinfo {author} {\bibfnamefont {Q.}~\bibnamefont {Gao}}, \bibinfo {author} {\bibfnamefont {B.}~\bibnamefont {Yao}}, \bibinfo {author} {\bibfnamefont {H.~S.}\ \bibnamefont {Pillai}}, \bibinfo {author} {\bibfnamefont {W.}~\bibnamefont {Zang}}, \bibinfo {author} {\bibfnamefont {X.}~\bibnamefont {Han}}, \bibinfo {author} {\bibfnamefont {Y.}~\bibnamefont {Liu}}, \bibinfo {author} {\bibfnamefont {S.-W.}\ \bibnamefont {Yu}}, \bibinfo {author} {\bibfnamefont {Z.}~\bibnamefont {Yan}}, \bibinfo {author} {\bibfnamefont {B.}~\bibnamefont {Min}}, \bibinfo {author} {\bibfnamefont {S.}~\bibnamefont {Zhang}},  \emph {et~al.},\ }\href@noop {} {\bibfield  {journal} {\bibinfo  {journal} {Nature Synthesis}\ }\textbf {\bibinfo {volume} {2}},\ \bibinfo {pages} {624} (\bibinfo {year} {2023})}\BibitemShut {NoStop}%
\bibitem [{\citenamefont {Darby}\ \emph {et~al.}(2018)\citenamefont {Darby}, \citenamefont {Sykes}, \citenamefont {Michaelides},\ and\ \citenamefont {Stamatakis}}]{darby2018carbon}%
  \BibitemOpen
  \bibfield  {author} {\bibinfo {author} {\bibfnamefont {M.~T.}\ \bibnamefont {Darby}}, \bibinfo {author} {\bibfnamefont {E.~C.~H.}\ \bibnamefont {Sykes}}, \bibinfo {author} {\bibfnamefont {A.}~\bibnamefont {Michaelides}}, \ and\ \bibinfo {author} {\bibfnamefont {M.}~\bibnamefont {Stamatakis}},\ }\href@noop {} {\bibfield  {journal} {\bibinfo  {journal} {Topics in catalysis}\ }\textbf {\bibinfo {volume} {61}},\ \bibinfo {pages} {428} (\bibinfo {year} {2018})}\BibitemShut {NoStop}%
\bibitem [{\citenamefont {Hannagan}\ \emph {et~al.}(2020)\citenamefont {Hannagan}, \citenamefont {Giannakakis}, \citenamefont {Flytzani-Stephanopoulos},\ and\ \citenamefont {Sykes}}]{hannagan2020single}%
  \BibitemOpen
  \bibfield  {author} {\bibinfo {author} {\bibfnamefont {R.~T.}\ \bibnamefont {Hannagan}}, \bibinfo {author} {\bibfnamefont {G.}~\bibnamefont {Giannakakis}}, \bibinfo {author} {\bibfnamefont {M.}~\bibnamefont {Flytzani-Stephanopoulos}}, \ and\ \bibinfo {author} {\bibfnamefont {E.~C.~H.}\ \bibnamefont {Sykes}},\ }\href@noop {} {\bibfield  {journal} {\bibinfo  {journal} {Chemical Reviews}\ }\textbf {\bibinfo {volume} {120}},\ \bibinfo {pages} {12044} (\bibinfo {year} {2020})}\BibitemShut {NoStop}%
\bibitem [{\citenamefont {McCue}\ and\ \citenamefont {Anderson}(2015)}]{mccue2015co}%
  \BibitemOpen
  \bibfield  {author} {\bibinfo {author} {\bibfnamefont {A.~J.}\ \bibnamefont {McCue}}\ and\ \bibinfo {author} {\bibfnamefont {J.~A.}\ \bibnamefont {Anderson}},\ }\href@noop {} {\bibfield  {journal} {\bibinfo  {journal} {Journal of Catalysis}\ }\textbf {\bibinfo {volume} {329}},\ \bibinfo {pages} {538} (\bibinfo {year} {2015})}\BibitemShut {NoStop}%
\bibitem [{\citenamefont {Ouyang}\ \emph {et~al.}(2021)\citenamefont {Ouyang}, \citenamefont {Papanikolaou}, \citenamefont {Boubnov}, \citenamefont {Hoffman}, \citenamefont {Giannakakis}, \citenamefont {Bare}, \citenamefont {Stamatakis}, \citenamefont {Flytzani-Stephanopoulos},\ and\ \citenamefont {Sykes}}]{ouyang2021directing}%
  \BibitemOpen
  \bibfield  {author} {\bibinfo {author} {\bibfnamefont {M.}~\bibnamefont {Ouyang}}, \bibinfo {author} {\bibfnamefont {K.~G.}\ \bibnamefont {Papanikolaou}}, \bibinfo {author} {\bibfnamefont {A.}~\bibnamefont {Boubnov}}, \bibinfo {author} {\bibfnamefont {A.~S.}\ \bibnamefont {Hoffman}}, \bibinfo {author} {\bibfnamefont {G.}~\bibnamefont {Giannakakis}}, \bibinfo {author} {\bibfnamefont {S.~R.}\ \bibnamefont {Bare}}, \bibinfo {author} {\bibfnamefont {M.}~\bibnamefont {Stamatakis}}, \bibinfo {author} {\bibfnamefont {M.}~\bibnamefont {Flytzani-Stephanopoulos}}, \ and\ \bibinfo {author} {\bibfnamefont {E.~C.~H.}\ \bibnamefont {Sykes}},\ }\href@noop {} {\bibfield  {journal} {\bibinfo  {journal} {Nature Communications}\ }\textbf {\bibinfo {volume} {12}},\ \bibinfo {pages} {1549} (\bibinfo {year} {2021})}\BibitemShut {NoStop}%
\bibitem [{\citenamefont {Xie}\ and\ \citenamefont {Grossman}(2018)}]{xie2018crystal}%
  \BibitemOpen
  \bibfield  {author} {\bibinfo {author} {\bibfnamefont {T.}~\bibnamefont {Xie}}\ and\ \bibinfo {author} {\bibfnamefont {J.~C.}\ \bibnamefont {Grossman}},\ }\href@noop {} {\bibfield  {journal} {\bibinfo  {journal} {Physical review letters}\ }\textbf {\bibinfo {volume} {120}},\ \bibinfo {pages} {145301} (\bibinfo {year} {2018})}\BibitemShut {NoStop}%
\bibitem [{\citenamefont {Hart}\ \emph {et~al.}(2021)\citenamefont {Hart}, \citenamefont {Mueller}, \citenamefont {Toher},\ and\ \citenamefont {Curtarolo}}]{hart2021machine}%
  \BibitemOpen
  \bibfield  {author} {\bibinfo {author} {\bibfnamefont {G.~L.}\ \bibnamefont {Hart}}, \bibinfo {author} {\bibfnamefont {T.}~\bibnamefont {Mueller}}, \bibinfo {author} {\bibfnamefont {C.}~\bibnamefont {Toher}}, \ and\ \bibinfo {author} {\bibfnamefont {S.}~\bibnamefont {Curtarolo}},\ }\href@noop {} {\bibfield  {journal} {\bibinfo  {journal} {Nature Reviews Materials}\ }\textbf {\bibinfo {volume} {6}},\ \bibinfo {pages} {730} (\bibinfo {year} {2021})}\BibitemShut {NoStop}%
\bibitem [{\citenamefont {Ubaru}\ \emph {et~al.}(2017)\citenamefont {Ubaru}, \citenamefont {Mi{\k{e}}dlar}, \citenamefont {Saad},\ and\ \citenamefont {Chelikowsky}}]{ubaru2017formation}%
  \BibitemOpen
  \bibfield  {author} {\bibinfo {author} {\bibfnamefont {S.}~\bibnamefont {Ubaru}}, \bibinfo {author} {\bibfnamefont {A.}~\bibnamefont {Mi{\k{e}}dlar}}, \bibinfo {author} {\bibfnamefont {Y.}~\bibnamefont {Saad}}, \ and\ \bibinfo {author} {\bibfnamefont {J.~R.}\ \bibnamefont {Chelikowsky}},\ }\href@noop {} {\bibfield  {journal} {\bibinfo  {journal} {Physical Review B}\ }\textbf {\bibinfo {volume} {95}},\ \bibinfo {pages} {214102} (\bibinfo {year} {2017})}\BibitemShut {NoStop}%
\bibitem [{\citenamefont {Hodges}\ \emph {et~al.}(1972)\citenamefont {Hodges}, \citenamefont {Watson},\ and\ \citenamefont {Ehrenreich}}]{hodges1972renormalized}%
  \BibitemOpen
  \bibfield  {author} {\bibinfo {author} {\bibfnamefont {L.}~\bibnamefont {Hodges}}, \bibinfo {author} {\bibfnamefont {R.}~\bibnamefont {Watson}}, \ and\ \bibinfo {author} {\bibfnamefont {H.}~\bibnamefont {Ehrenreich}},\ }\href@noop {} {\bibfield  {journal} {\bibinfo  {journal} {Physical Review B}\ }\textbf {\bibinfo {volume} {5}},\ \bibinfo {pages} {3953} (\bibinfo {year} {1972})}\BibitemShut {NoStop}%
\bibitem [{\citenamefont {Gelatt~Jr}\ \emph {et~al.}(1977)\citenamefont {Gelatt~Jr}, \citenamefont {Ehrenreich},\ and\ \citenamefont {Watson}}]{gelatt1977renormalized}%
  \BibitemOpen
  \bibfield  {author} {\bibinfo {author} {\bibfnamefont {C.}~\bibnamefont {Gelatt~Jr}}, \bibinfo {author} {\bibfnamefont {H.}~\bibnamefont {Ehrenreich}}, \ and\ \bibinfo {author} {\bibfnamefont {R.}~\bibnamefont {Watson}},\ }\href@noop {} {\bibfield  {journal} {\bibinfo  {journal} {Physical Review B}\ }\textbf {\bibinfo {volume} {15}},\ \bibinfo {pages} {1613} (\bibinfo {year} {1977})}\BibitemShut {NoStop}%
\bibitem [{\citenamefont {Ralchenko}(2005)}]{ralchenko2005nist}%
  \BibitemOpen
  \bibfield  {author} {\bibinfo {author} {\bibfnamefont {Y.}~\bibnamefont {Ralchenko}},\ }\href@noop {} {\bibfield  {journal} {\bibinfo  {journal} {Memorie della Societa Astronomica Italiana Supplementi}\ }\textbf {\bibinfo {volume} {8}},\ \bibinfo {pages} {96} (\bibinfo {year} {2005})}\BibitemShut {NoStop}%
\bibitem [{\citenamefont {Hammer}\ and\ \citenamefont {N{\o}rskov}(1997)}]{hammer1997theory}%
  \BibitemOpen
  \bibfield  {author} {\bibinfo {author} {\bibfnamefont {B.}~\bibnamefont {Hammer}}\ and\ \bibinfo {author} {\bibfnamefont {J.~K.}\ \bibnamefont {N{\o}rskov}},\ }in\ \href@noop {} {\emph {\bibinfo {booktitle} {Chemisorption and Reactivity on Supported Clusters and Thin Films: Towards an Understanding of Microscopic Processes in Catalysis}}}\ (\bibinfo  {publisher} {Springer},\ \bibinfo {year} {1997})\ pp.\ \bibinfo {pages} {285--351}\BibitemShut {NoStop}%
\bibitem [{\citenamefont {Brooks}\ and\ \citenamefont {Johansson}(1983)}]{brooks1983exchange}%
  \BibitemOpen
  \bibfield  {author} {\bibinfo {author} {\bibfnamefont {M.}~\bibnamefont {Brooks}}\ and\ \bibinfo {author} {\bibfnamefont {B.}~\bibnamefont {Johansson}},\ }\href@noop {} {\bibfield  {journal} {\bibinfo  {journal} {Journal of Physics F: Metal Physics}\ }\textbf {\bibinfo {volume} {13}},\ \bibinfo {pages} {L197} (\bibinfo {year} {1983})}\BibitemShut {NoStop}%
\bibitem [{\citenamefont {Huang}\ \emph {et~al.}(2024{\natexlab{b}})\citenamefont {Huang}, \citenamefont {Wang}, \citenamefont {Kamanuru}, \citenamefont {Achenie}, \citenamefont {Kitchin},\ and\ \citenamefont {Xin}}]{huang2024unifying}%
  \BibitemOpen
  \bibfield  {author} {\bibinfo {author} {\bibfnamefont {Y.}~\bibnamefont {Huang}}, \bibinfo {author} {\bibfnamefont {S.-H.}\ \bibnamefont {Wang}}, \bibinfo {author} {\bibfnamefont {M.}~\bibnamefont {Kamanuru}}, \bibinfo {author} {\bibfnamefont {L.~E.}\ \bibnamefont {Achenie}}, \bibinfo {author} {\bibfnamefont {J.~R.}\ \bibnamefont {Kitchin}}, \ and\ \bibinfo {author} {\bibfnamefont {H.}~\bibnamefont {Xin}},\ }\href@noop {} {\bibfield  {journal} {\bibinfo  {journal} {Physical Review B}\ }\textbf {\bibinfo {volume} {110}},\ \bibinfo {pages} {L121404} (\bibinfo {year} {2024}{\natexlab{b}})}\BibitemShut {NoStop}%
\bibitem [{\citenamefont {Kitchin}\ \emph {et~al.}(2004{\natexlab{b}})\citenamefont {Kitchin}, \citenamefont {N{\o}rskov}, \citenamefont {Barteau},\ and\ \citenamefont {Chen}}]{kitchin2004role}%
  \BibitemOpen
  \bibfield  {author} {\bibinfo {author} {\bibfnamefont {J.~R.}\ \bibnamefont {Kitchin}}, \bibinfo {author} {\bibfnamefont {J.~K.}\ \bibnamefont {N{\o}rskov}}, \bibinfo {author} {\bibfnamefont {M.~A.}\ \bibnamefont {Barteau}}, \ and\ \bibinfo {author} {\bibfnamefont {J.}~\bibnamefont {Chen}},\ }\href@noop {} {\bibfield  {journal} {\bibinfo  {journal} {Physical review letters}\ }\textbf {\bibinfo {volume} {93}},\ \bibinfo {pages} {156801} (\bibinfo {year} {2004}{\natexlab{b}})}\BibitemShut {NoStop}%
\bibitem [{\citenamefont {Clementi}\ and\ \citenamefont {Raimondi}(1963)}]{clementi1963atomic}%
  \BibitemOpen
  \bibfield  {author} {\bibinfo {author} {\bibfnamefont {E.}~\bibnamefont {Clementi}}\ and\ \bibinfo {author} {\bibfnamefont {D.-L.}\ \bibnamefont {Raimondi}},\ }\href@noop {} {\bibfield  {journal} {\bibinfo  {journal} {The Journal of Chemical Physics}\ }\textbf {\bibinfo {volume} {38}},\ \bibinfo {pages} {2686} (\bibinfo {year} {1963})}\BibitemShut {NoStop}%
\bibitem [{\citenamefont {Bueno}\ \emph {et~al.}(2022)\citenamefont {Bueno}, \citenamefont {Leonardi}, \citenamefont {Kar}, \citenamefont {Chatterjee}, \citenamefont {Zhan}, \citenamefont {Chen}, \citenamefont {Wang}, \citenamefont {Engel}, \citenamefont {Fung},\ and\ \citenamefont {Skrabalak}}]{bueno2022quinary}%
  \BibitemOpen
  \bibfield  {author} {\bibinfo {author} {\bibfnamefont {S.~L.}\ \bibnamefont {Bueno}}, \bibinfo {author} {\bibfnamefont {A.}~\bibnamefont {Leonardi}}, \bibinfo {author} {\bibfnamefont {N.}~\bibnamefont {Kar}}, \bibinfo {author} {\bibfnamefont {K.}~\bibnamefont {Chatterjee}}, \bibinfo {author} {\bibfnamefont {X.}~\bibnamefont {Zhan}}, \bibinfo {author} {\bibfnamefont {C.}~\bibnamefont {Chen}}, \bibinfo {author} {\bibfnamefont {Z.}~\bibnamefont {Wang}}, \bibinfo {author} {\bibfnamefont {M.}~\bibnamefont {Engel}}, \bibinfo {author} {\bibfnamefont {V.}~\bibnamefont {Fung}}, \ and\ \bibinfo {author} {\bibfnamefont {S.~E.}\ \bibnamefont {Skrabalak}},\ }\href@noop {} {\bibfield  {journal} {\bibinfo  {journal} {ACS nano}\ }\textbf {\bibinfo {volume} {16}},\ \bibinfo {pages} {18873} (\bibinfo {year} {2022})}\BibitemShut {NoStop}%
\bibitem [{\citenamefont {Huang}\ \emph {et~al.}(2024{\natexlab{c}})\citenamefont {Huang}, \citenamefont {Wang}, \citenamefont {Wang}, \citenamefont {Omidvar}, \citenamefont {Achenie}, \citenamefont {Skrabalak},\ and\ \citenamefont {Xin}}]{huang2024unraveling}%
  \BibitemOpen
  \bibfield  {author} {\bibinfo {author} {\bibfnamefont {Y.}~\bibnamefont {Huang}}, \bibinfo {author} {\bibfnamefont {S.-H.}\ \bibnamefont {Wang}}, \bibinfo {author} {\bibfnamefont {X.}~\bibnamefont {Wang}}, \bibinfo {author} {\bibfnamefont {N.}~\bibnamefont {Omidvar}}, \bibinfo {author} {\bibfnamefont {L.~E.}\ \bibnamefont {Achenie}}, \bibinfo {author} {\bibfnamefont {S.~E.}\ \bibnamefont {Skrabalak}}, \ and\ \bibinfo {author} {\bibfnamefont {H.}~\bibnamefont {Xin}},\ }\href@noop {} {\bibfield  {journal} {\bibinfo  {journal} {The Journal of Physical Chemistry C}\ }\textbf {\bibinfo {volume} {128}},\ \bibinfo {pages} {11183} (\bibinfo {year} {2024}{\natexlab{c}})}\BibitemShut {NoStop}%
\bibitem [{\citenamefont {Kress}\ \emph {et~al.}(2023)\citenamefont {Kress}, \citenamefont {Zhang}, \citenamefont {Wang}, \citenamefont {{\c{C}}{\i}nar}, \citenamefont {Friend}, \citenamefont {Sykes},\ and\ \citenamefont {Montemore}}]{kress2023priori}%
  \BibitemOpen
  \bibfield  {author} {\bibinfo {author} {\bibfnamefont {P.~L.}\ \bibnamefont {Kress}}, \bibinfo {author} {\bibfnamefont {S.}~\bibnamefont {Zhang}}, \bibinfo {author} {\bibfnamefont {Y.}~\bibnamefont {Wang}}, \bibinfo {author} {\bibfnamefont {V.}~\bibnamefont {{\c{C}}{\i}nar}}, \bibinfo {author} {\bibfnamefont {C.~M.}\ \bibnamefont {Friend}}, \bibinfo {author} {\bibfnamefont {E.~C.~H.}\ \bibnamefont {Sykes}}, \ and\ \bibinfo {author} {\bibfnamefont {M.~M.}\ \bibnamefont {Montemore}},\ }\href@noop {} {\bibfield  {journal} {\bibinfo  {journal} {Journal of the American Chemical Society}\ }\textbf {\bibinfo {volume} {145}},\ \bibinfo {pages} {8401} (\bibinfo {year} {2023})}\BibitemShut {NoStop}%
\end{thebibliography}%

\vskip 6pt
\noindent 
All source data along with Jupyter notebooks for data preprocessing, model development, and \emph{post hoc} analysis are available from the GitHub repository: \url{https://github.com/hlxin/XXXXXXXXXXXXXXXXXXXXXXXXXXX}. 
~\

\vskip 6pt
\noindent Y.H. and H.X. acknowledge the financial support from the US Department of Energy, Office of Basic Energy Sciences under contract  no.~DE-SC0023323. S.H.W. and L.E.K.A. especially thank to the NSF Non-Academic Research Internships for Graduate Students (INTERN) program (CBET-1845531) for supporting part of the work in NIST under the guidance of Dr.~Kamal Choudhary. The computational resource used in this work is provided by the advanced research computing at Virginia Polytechnic Institute and State University. 

~\

\vskip 6pt
\noindent H.X. and Y.H. conceived the idea and designed the computational approach. Y.H. and S.H.W. conducted DFT calculations and machine learning model development. Y.H. performed the detailed model interpretation. All authors contributed to the writing and editing of this manuscript.

~\

\vskip 6pt
\noindent The authors declare no competing interests.

\newcommand{\mycomment}[1]{}

\mycomment{

}

\message{The column width is: \the\columnwidth}

\end{document}